# SMAP-BASED RETRIEVAL OF VEGETATION OPACITY AND ALBEDO

*Dara Entekhabi[1], Alexandra Konings[2], Maria Piles[3] and Narendra Das[4]*

[1]Department of Civil and Environmental Engineering, Massachusetts Institute of Technology, Cambridge 02139, USA
[2]Department of Earth System Science, Stanford University, Stanford, CA
[3]Remote Sensing Laboratory, Departament de Teoria del Senyal i Comunicacions, Universitat Politecnica de Catalunya (UPC), 08034 Barcelona, Spain
[4]Jet Propulsion Laboratory, California Institute of Technology, Pasadena, CA , 91109

## ABSTRACT

Over land the vegetation canopy affects the microwave brightness temperature by emission, scattering and attenuation of surface soil emission.  The questions addressed in this study are: 1) what is the transparency of the vegetation canopy for different biomes around the Globe at the low-frequency L-band?, 2) what is the seasonal amplitude of vegetation microwave optical depth for different biomes?, 3) what is the effective scattering at this frequency for different vegetation types?, 4) what is the impact of imprecise characterization of vegetation microwave properties on retrieval of soil surface conditions? These questions are addressed based on the recently completed one full annual cycle measurements by the NASA Soil Moisture Active Passive (SMAP) measurements.

## 1. INTRODUCTION

For monitoring vegetation abundance and function, microwave radiometry offers distinct advantages over optical indices because the atmosphere is nearly transparent (sensing regardless of cloud cover and solar illumination) and the canopy may be partially penetrated with microwaves (Shi et al., 2013; Du et al., 2016). In this study we extend the applicability of the new global L-band radiometry to the terrestrial biosphere. We propose that the measurements of the surface at this microwave frequency are as valuable to global ecology as they have proven to be for the ocean and hydrologic sciences.  In this study we use the one full annual cycle of SMAP radiometer data to derive information of the water status and structure of vegetation canopy around the globe. The new information is complementary to existing visible/infrared (e.g., vegetation indices such as Extended Vegetation Index [EVI], Normalized Difference Vegetation Index [NDVI], Solar-Induced Fluorescence [SIF]), laser altimeter tree height and radar scattering vegetation indicators.  The study shows that useful information on the canopy status can be derived from the L-band radiometer measurements. Importantly the process of extracting the information does not rely on any ancillary information that could influence the spatial and temporal patterns of the vegetation indicators. As independent information on the vegetation canopy, the new data complements the visible/infrared, lidar and radar data. Together they can enable new capabilities in monitoring ecosystem function in different biomes (e.g., water relations in the soil-vegetation continuum and ecosystem response to dry episodes and seasons) and potentially reveal evolutionary adaptation of different biomes and vegetation types to water stress, photosynthetic radiation availability and nutrient, predator and competition stresses.

## 2. APPROACH

In this study we implement an interpretation framework that uses two-channels of the SMAP radiometer (H- and V-pol) to infer the three key variables ($\tau$, $\omega$ and $k$) without reliance on ancillary information on vegetation. The inputs to the algorithm are the two SMAP brightness temperature measurements and an estimate of physical temperature at 06:00 local time. The outputs are $\tau$, $\omega$ and soil rough surface reflectivity $r_p$. The approach recognizes that the number of unknowns (in this case three) is less than the number of measurements (two and less owing to the correlation among the observations). In this study we estimate the number of independent information (fractional between one and two) for the two SMAP measurements using the Degrees-of-Information (DoI) introduced by Konings et. al. (2015). Based on the DoI of observations, we implement a retrieval algorithm for $\tau$, $\omega$ and $r_p$. The approach is to use adjacent-in-time overpasses (two to three days apart) to infer faster-changing soil and slower-changing vegetation characteristics.

The attempt to retrieve more parameters than the DoI of the observations can result in multitude of local minima in the retrieval cost function. Under these circumstances the retrievals are susceptible to observations noise and initial conditions. Attempts to retrieve more parameters than the DoI is prevalent in the case of tau-omega-based surface soil moisture and vegetation microwave opacity retrievals. These approaches are often characterized by higher noise standard deviation owing to jumps between local minima. They are often also characterized by high incidence of non-convergence. It is important to measure the DoI of observations and avoid retrieval of more parameters than information in the observations allow.

The estimates of $\tau$ and $\omega$ in this case are based on the low-frequency microwave measurements and free of any influence from visible/infrared vegetation indices or classifications. As such they serve the goals of this study which is to introduce new/complementary information on the dynamic global terrestrial biosphere in the recently emerging new era of global low-frequency microwave remote sensing.

The new dynamic maps of $\tau$ and $\omega$ should also help reduce the uncertainty of estimates used in the SMAP surface soil moisture retrieval algorithms. Again these algorithms rely on NDVI seasonal climatology and static vegetation visible/infrared-based vegetation classification.

## 3. DATA AND ANALYSES

The L-band brightness temperature (horizontal and vertical polarization) used in this study are from the enhanced Chaubell et al. (2016) enhanced SMAP radiometer product. The period of coverage is one full annual cycle spanning April 1, 2015 to March 31, 2016. The data is posted at 9 [km] over an Equal-Area Scalable Earth-2 (EASE2) grid domain. The resolution of the brightness temperatures are about 40 [km] based on the geometric mean of the major and minor axes defined by the oval containing half of the power (-3 [dB]) across the SMAP antenna gain.

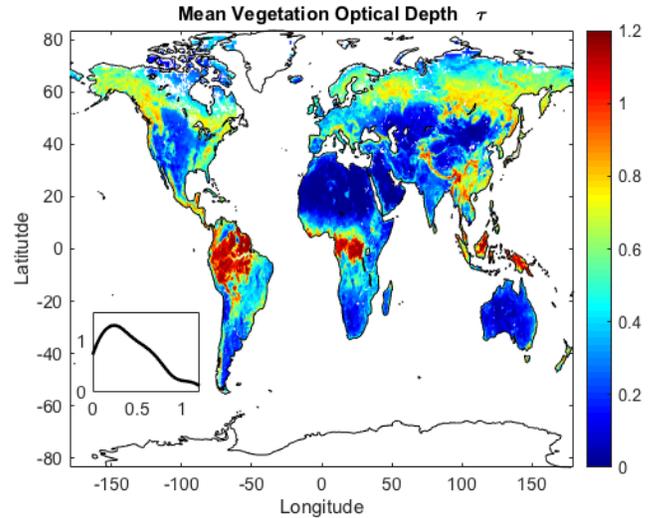

**Fig. 1**. Global distribution of the time-averaged vegetation optical depth $\tau$ at nadir (dimensionless). Retrievals are based on the application of the dual-channel time-series algorithm on one year of SMAP radiometer instrument measurements. The maximum values are concentrated across dense and wet tropical forests and correspond to transmissivity values as low as 15% at 40 degrees. Seasonally dry tropics and savannas (central Africa and southern Brazil for example) are distinct from the nearby wet tropics. Boreal forests have high values as well reaching transmissivity values up to 40%. The inset figure in this and subsequent maps are an estimate of the mapped data marginal probability density.

## 4. RESULTS

The results of our study and assessment of the four questions are: The L-band brightness temperature-based retrievals of the vegetation optical depth follow closely the height of vegetation (based on comparisons with independent lidar measurements). The median extinction (at 40 [deg]) due to the vegetation canopy is about 23% and can reach upto 60%. The amplitude of the microwave optical depth which is positively related to the wet biomass content of the canopy follows patterns of light-limitation in vegetation growth. The effective single-scattering globally has a median close to 8% and half of the retrieved values are within 5% of the median.

The results confirm and expand on the more limited (in spatial and temporal resolution) study by Konings et al. (2016) using Aquarius measurements. The approach of this study also allows estimation of a dynamic ω as well as dynamic τ.

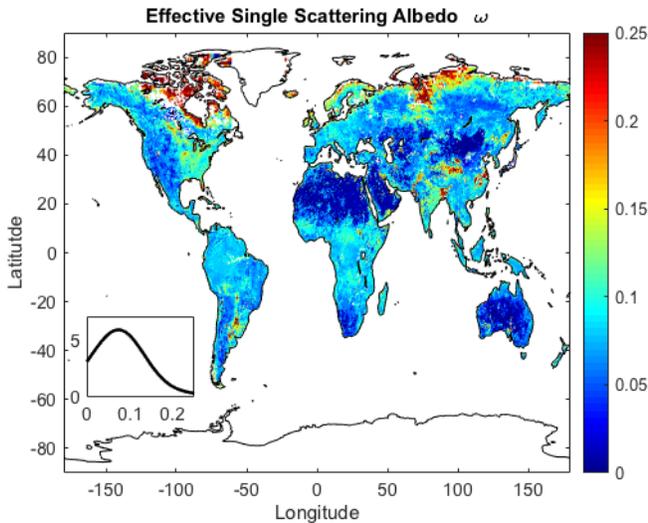

**Fig. 2**. Retrievals of the effective single-scattering albedo ω (dimensionless) based on one year of SMAP radiometer measurements and the multi-temporal dual-channel algorithm. The values are generally below 12% for all vegetation types. Regions with values as high as 25% are restricted to far northern boreal regions. These regions contain a high density of small lakes and inland water bodies. Inadequate correction of water-body contribution in the brightness temperature data can result in low brightness temperatures. An inflated vegetation albedo can fit the brightness temperature values. In other regions of the globe the effective single-scattering albedo ω is generally between 0% and 12% with the more densely vegetated areas (eastern North America and tropical Africa for example) showing values in the upper half of the range. Sparsely-vegetated areas (western US, subtropical Africa, central Asia and Australia for example) show with values in the 0% to 5% range.

The dynamic fields of τ and ω are compatible (in magnitude and regional patterns) with the retrievals based on ESA Soil Moisture and Salinity (SMOS) retrievals which are also L-band brightness temperature measurements but are multi-angular. The angular information allows retrieval of vegetation opacity and surface reflectivity simultaneously. In forthcoming studies regional perspectives are used to focus on particular biomes (e.g., agro-ecosystems, wet and dry tropical forests). The regional studies use knowledge of the dominant vegetation phenology to select a relevant window for the estimation of dynamic vegetation parameters. With radiometry-based retrieval of surface volumetric soil water content and vegetation microwave optical depth, the studies diagnose timing of soil water uptake in the soil-vegetation continuum as well as characterize the vegetation growth response to seasonal water- and light-limitation.

## 4. REFERENCES

Chaubell, M. J., S. Chan, R. S. Dunbar, J. Peng, and S. Yueh. 2016. *SMAP Enhanced L1C Radiometer Half-Orbit 9 km EASE-Grid Brightness Temperatures, Version 1*. [Indicate subset used]. Boulder, Colorado USA. NASA National Snow and Ice Data Center Distributed Active Archive Center. doi: http://dx.doi.org/10.5067/2C9O9KT6JAWS.

Du, J., Kimball, J. S., & Jones, L. A. (2016). Passive Microwave Remote Sensing of Soil Moisture Based on Dynamic Vegetation Scattering Properties for AMSR-E. IEEE Transactions on Geoscience and Remote Sensing, 54(1), 597-608.

Konings, A. G., McColl, K. A., Piles, M., & Entekhabi, D. (2015). How many parameters can be maximally estimated from a set of measurements?. *IEEE Geoscience and Remote Sensing Letters*, *12*(5), 1081-1085.

Konings, A. G., Piles, M., Rötzer, K., McColl, K. A., Chan, S. K., & Entekhabi, D. (2016). Vegetation optical depth and scattering albedo retrieval usingtime series of dual-polarized L-band radiometer observations. *Remote sensing of environment*, *172*, 178-189.

Shi, J., Jackson, T., Tao, J., Du, J., Bindlish, R., Lu, L., & Chen, K. S. (2008). Microwave vegetation indices for short vegetation covers from satellite passive microwave sensor AMSR-E. Remote Sensing of Environment, 112(12), 4285-4300. Kurum, Mehmet. "Quantifying scattering albedo in microwave emission of vegetated terrain." *Remote Sensing of Environment* 129 (2013): 66-74.